\documentclass[12pt,prd,aps,amssymb,amsmath,tightenlines,showpacs]{article}
\usepackage[utf8]{inputenc}
\usepackage[a4paper,
top=3cm,bottom=3cm,left=1.5cm,right=1.5cm]{geometry}
\usepackage{amssymb}
\usepackage{amsmath}
\usepackage{amsthm}
\usepackage{tipa}
\usepackage{latexsym} 
\usepackage{graphicx}
\usepackage{slashed}
\usepackage{bbold}
\usepackage{cancel}
\usepackage{comment}
\usepackage{physics}
\usepackage{color}
\usepackage{graphicx,epsfig,color}
\usepackage{comment}
\usepackage{cite}
\usepackage{tikz}
\usepackage[compat=1.1.0]{tikz-feynman}
\usetikzlibrary{decorations.markings}
\usepackage{color}
\usepackage{graphicx,epsfig,color}
\usepackage[dvipsnames]{xcolor}
\usepackage{hyperref}
\hypersetup{colorlinks, linkcolor=Magenta,citecolor=Cerulean
}

\tikzfeynmanset{
	photon to bigotimes/.style={
		/tikz/postaction={
			decorate,
			decoration={
				markings,
				mark=at position 0.84 with {
					\node[anchor=west] {$\bigotimes$};
				}
			}
		}
	}
}

\tikzfeynmanset{
	photon to bigotimes2/.style={
		/tikz/postaction={
			decorate,
			decoration={
				markings,
				mark=at position 0.80 with {
					\node[anchor=west] {$\bigotimes$};
				}
			}
		}
	}
}

\newcommand{\be}{\begin{equation}}
\newcommand{\ee}{\end{equation}}
\newcommand{\bea}{\begin{eqnarray}}
\newcommand{\eea}{\end{eqnarray}}
\newcommand{\vv}{``}

\def\nn{\nonumber} 
\numberwithin{equation}{section}

\begin{document}
	\graphicspath{{FIGURE/}}
	\topmargin=-2cm

\begin{center} {\Large{\bf Hard cutoff and gauge theories}}

\vspace*{0.8 cm}

	{\small 
V. Branchina$^{\dagger,\,a,b},$ \let\thefootnote\relax\footnote{$^\dagger$branchina@ct.infn.it}F. Contino$^{\ddagger,\,c,d},$ \footnote{$^\ddagger$f.contino@ssmeridionale.it}\,R. Gandolfo$^{\star,\,a,b},$ \footnote{$^\star$riccardo.gandolfo@ct.infn.it }A. Pernace$^{*,\,a,b}$ \footnote{$^*$arcangelo.pernace@ct.infn.it}}
		
				\vspace*{0.4cm}
		
	{\it	${}^a${\footnotesize Department of Physics, University of Catania, Via Santa Sofia 64, I-95123 Catania, Italy}
		
		\vskip 5pt
		
			${}^b${\footnotesize INFN-Catania,
						Via Santa Sofia 64, I-95123 
						Catania, Italy}
				
				\vskip 5pt
				
				${}^c${\footnotesize Scuola Superiore Meridionale, Largo San Marcellino 10, 80138 Napoli, Italy}
				
				\vskip 5pt
				
				${}^d${\footnotesize INFN-Napoli, Complesso Universitario di Monte S. Angelo, Via Cinthia Edificio 6, 80126 Napoli, Italy}}
		
\vskip 1cm

\setcounter{footnote}{0}

{\bf\small Abstract}
\end{center}

{\small
\noindent
According to usual calculations, the use of a hard cutoff $\Lambda$ in gauge theories leads to a violation of gauge invariance. This seems to generate a tension between gauge theories and the Wilsonian effective field theory (EFT) paradigm, where $\Lambda$ has the physical meaning of ultimate scale of the theory, the scale above which the latter has to be replaced by its UV completion. In the present work, considering the Euler-Heisenberg correction to the free Maxwell action, we present a way to introduce the Wilsonian hard UV cutoff $\Lambda$ that preserves gauge invariance at the quantum level. For both scalar and fermionic QED, we recover the well-known Euler-Heisenberg result obtained within proper-time regularization, apart from terms that are generically cutoff-suppressed. These terms, periodic in the inverse background field, might become relevant in regimes where the latter probes scales not much smaller than $\Lambda$. On the theoretical side, the methods developed in the present work represent a first step towards a new (closer in spirit to the Wegner-Houghton construction) realization of the Wilsonian renormalization group program in gauge theories.}

\section{Introduction}

As it is well-known, a great progress in our understanding of renormalization in quantum field theory (QFT)
comes from Wilson's lesson \cite{Wilson:1971bg,Wilson:1971dc,Wilson:1971dh,Wilson:1973jj, Wegner:1972ih,Wilson:1974mb}, according to which any QFT is an effective field theory (EFT), i.e.\,\,a theory defined with a built-in UV  physical cut-off $\Lambda$, the scale above which it is no longer valid and has to be replaced by its UV-completion. The corresponding Lagrangian 
${\mathcal L}^{(\Lambda)}$
allows to describe processes at energies $E\leq\Lambda$, and the \vv bare'' parameters $g_i(\Lambda)$ in ${\mathcal L}^{(\Lambda)}$ keep track of the higher energy degrees of freedom ($E>\Lambda$) of the UV completion of the theory. 
A fundamental aspect of the Wilsonian paradigm consists in including in the effective action the quantum fluctuations through the progressive integration of modes in infinitesimal shells, starting from the UV scale $\Lambda$ and moving towards the IR. At energy scales $\mu<\Lambda$, the parameters $g_i (\mu)$ result from the elimination of the modes in the range $[\mu, \Lambda]$, and this gives rise to the renormalization group (RG) flow of the theory. 

According to usual calculations, the use of a hard cutoff $\Lambda$ in gauge theories leads to a violation of the Ward-Takahashi (Slavnov-Taylor) identities \cite{Ward:1950xp,Takahashi:1957xn,Taylor:1971ff,Slavnov:1972fg}. A simple example is given by the vacuum polarization tensor $\Pi_{\mu\nu}$ in QED. At one-loop order, a quadratically divergent longitudinal contribution arises, that can be traced back to a non-gauge invariant term $\sim \Lambda^2 A_\mu A^\mu$ in the effective action. Faced with this problem, the common attitude is to abandon the idea of a hard cutoff and to resort to regularization methods that by construction preserve gauge invariance. Pauli-Villars, dimensional and zeta-function regularization, proper-time are among the most used ones.
The physical meaning of these methods, however, is less transparent. Pauli and Villars themselves acknowledged that their method, based on the introduction of  fictitious ghost-like particles with heavy masses $M_i$, had to be regarded as a \vv formalistic'' \vv transitional stage of the theory'', awaiting a truly physical way to preserve gauge invariance at the quantum level \cite{Pauli:1949zm}. Analogous considerations hold for dimensional and zeta-function regularization, where non-gauge invariant terms are canceled due to the use of complex analysis techniques \cite{Branchina:2022jqc,Branchina:2022gll}. 

According to the above considerations, there seems to be a tension between the Wilsonian paradigm and gauge theories. In view of the deep physical understanding brought by Wilson's lesson, and of the central role played by gauge theories in the description of fundamental interactions, it seems of great importance to find a way to reconcile these two fundamental pillars of modern physics. Some attempts in this direction are in \cite{Bonini:1993kt,Bonini:1993sj,Bonini:1994kp,Bonini:1994dz,DAttanasio:1996tzp,Bonini:1996bk,Simionato:1998te,Simionato:1998iz,Morris:1999px,Simionato:2000ut,Morris:2000fs,Gies:2001nw,Gies:2002af,Arnone:2002cs, Gies:2004hy,Gies:2005as,Arnone:2005vd,Pawlowski:2005xe,Morris:2006in,Arnone:2005fb,Schaefer:2007pw,Fischer:2008uz,Braun:2007bx,Gies:2006wv,Fu:2019hdw,Dupuis:2020fhh,Branchina:2023ogv,Branchina:2024ljd,Branchina:2024xzh,Branchina:2024lai,Branchina:2025lqw,Branchina:2025hen,Giacometti:2025qyy,Giacometti:2026zrs}.

In the present work, we investigate the possibility of introducing a Wilsonian {\it hard} \,UV physical cutoff $\Lambda$ that preserves gauge invariance. To this end, it is sufficient to consider the simple case\footnote{The case of non-Abelian gauge theories will be considered elsewhere \cite{us}.} of the Euler-Heisenberg correction $\delta S_{_{\rm EH}}$ \cite{Heisenberg:1936nmg, Schwinger:1951nm,Dittrich:1975au, Dunne:2004nc, Dunne:2012vv, Weisskopf:1936hya, Gies:2016yaa} to the free Maxwell action $S_{\rm em}=-\frac14\int\dd^4x F_{\mu\nu}F^{\mu\nu}$. We will perform the calculation both in the case of scalar and fermionic QED.
Following the suggestion coming from Wilson's lesson, in the present work we calculate the fluctuation determinant that appears in \(\delta S_{_{\rm EH}}\) as the product over the eigenvalues
\(\lambda_n\) of the fluctuation operator, and implement the Wilsonian UV physical cutoff \(\Lambda\) as a hard cut over the
spectrum: \(\lambda_n\leq\Lambda^2\). We will consider the case of a constant magnetic field background, in which case the spectrum is given by the Landau levels. As we will see, our result coincides with the well-known
Schwinger's result \cite{Schwinger:1951nm} (obtained with proper-time regularization) apart
from the appearance of field-dependent oscillatory terms. These terms are generically suppressed
by inverse powers of the UV cutoff \(\Lambda\), but may become relevant when the background field probes scales
not much smaller than $\Lambda$.

Oscillatory contributions originating from the discreteness
of Landau levels and from the presence of a sharp physical boundary are well known in condensed
matter physics. The paradigmatic example is provided by de Haas-van Alphen oscillations, where
the magnetization displays a periodic dependence on the inverse magnetic field as
Landau levels cross the Fermi surface \cite{deHaasvanAlphen:1930, Shoenberg:1984}. Related quantum oscillatory phenomena
also occur in Dirac and Weyl materials, in moir\'e graphene systems, and in Hofstadter-type spectra
realized in graphene-based superlattices and cold-atom optical lattices
\cite{Hofstadter:1976,Janecek:2013,Dean:2013, Hunt:2013, Aidelsburger:2013, Bocarsly:2024}. Of course, the physical setting is
different from the one considered here: in those systems the sharp boundary is typically provided
by a Fermi surface, a band edge or a lattice miniband. In the present work, instead, the sharp boundary is the Wilsonian
spectral cutoff. To the best of our knowledge, this is the first time that
oscillatory terms of this kind are discussed 
in the context of fundamental gauge theories.

The rest of the paper is organized as follows. In section \ref{EH-scalar}, we recall the definition of the Euler-Heisenberg correction $\delta S^\phi_{_{\rm EH}}$ to the Maxwell action in the case of scalar QED and set up the tools for our analysis, paying attention to the scale $M$ that takes care of the dimensions of the fluctuation operator. In section \ref{eig-scalar}, we solve the eigenvalue problem for the (Euclidean) fluctuation operator specifying to the case of a constant magnetic background $B={\rm const}$, taking $E=0$, and calculate the corresponding Euler-Heisenberg correction $\delta S^\phi_{_{\rm EH}}$. In section \ref{EH-fermionic}, we perform the same calculation for the case of fermionic QED. In section \ref{discussion}, we further discuss the issue of the scale $M$ used in the calculation of the fluctuation determinant, considering also the relation between our calculation (spectral hard cutoff) and proper-time regularization. Section \ref{Conclusions} is for conclusions and outlooks.

\section{Euler-Heisenberg action. Scalar QED}
\label{EH-scalar}

In this section we perform the calculation of the Euler-Heisenberg action introducing the Wilsonian UV physical cutoff $\Lambda$. Before moving to that, it is worth to discuss an important issue related to the path integral formulation of QFTs. To illustrate the point, we consider the simple theory of a single component real scalar field (Euclidean signature),
\begin{equation}\label{phi4}
	\mathcal L = \frac 12 \partial_\mu \phi\, \partial^\mu \phi  + \frac 12 m^2 \phi^2 + \frac{\lambda}{4!} \phi^4 \,.
\end{equation}
For a constant background $\Phi$, the one-loop correction to the classical potential is usually obtained directly from the configuration space path integral $\int \mathcal D \phi\, e^{-S[\phi]}$ and is written as
\begin{equation}\label{deltaV}
	\delta V_{\rm 1l} (\Phi) = \Tr \log \frac{-\partial_\mu \partial^\mu + m^2 + \frac{\lambda}{2} \Phi^2}{M^2}\,,
\end{equation} 
where $M$ is a mass scale that is typically introduced to take care of the dimensions inside the logarithm. Here we stress an important point. Though after calculation of the trace in\,\eqref{deltaV} the scale $M$ appears in $\delta V_{\rm 1l} (\Phi)$ only in unimportant field-independent terms (vacuum energy), and there is then no need to specify this scale, there might be situations where this is not the case. As we will see, this is what happens in the case of the Euler-Heisenberg Lagrangian. These issues, and their connection with proper-time regularization, will be further discussed in section \ref{discussion}. 

To find which scale takes care of the dimensions in \eqref{deltaV}, it is useful to refer to the original phase space path integral of the theory\,\eqref{phi4}. Let us consider the corresponding Hamiltonian density $\mathcal H(\pi,\phi) = \frac{1}{2} \pi^2 + \frac{1}{2} |\nabla \phi|^2 + \frac 12 m^2 \phi^2 + \frac{\lambda}{4!}\phi^4$ (quadratic in the conjugate momenta $\pi$). The vacuum persistence amplitude is
\begin{align}\label{phasespace}
	\int\text{\small$\Big(\prod_x\frac{a^3\dd\pi_x}{2\pi}\Big)\Big(\prod_x\dd\phi_x\Big)$} \exp \text{\small$\Big(\sum_x a^4\,\Big[\,i \pi_x\partial_0\phi-\frac{1}{2}\pi_x^2-\frac{1}{2} |\nabla \phi|^2 - \frac 12 m^2 \phi^2 - \frac{\lambda}{4!}\phi^4\,\Big]\Big)$}\,\,,
\end{align}

\noindent
where {\small $\prod_{x}$} indicates the product over the points of a 4D grid of lattice spacing $a$, and $\partial_0\phi$ and $\nabla \phi$ indicate the discretized version of the (Euclidean) time derivative and of the gradient, respectively. The integration over the conjugate momenta $\pi_x$ leads to
\begin{align}\label{dimless}
	\int\text{\small$\prod_x  \dd (\frac{a \phi_x}{\sqrt{2 \pi}})$} e^{-\sum_x a^4\,\mathcal L (\phi_x,\partial_\mu \phi)}\,\,,
\end{align}
where $\mathcal L$ is the (discretized version of the) Lagrangian density\,\eqref{phi4}, and \,$a \phi_x$\, are the dimensionless integration variables. The resulting configuration space path integral measure  {\small $\prod_x  \dd (a \phi_x/\sqrt{2 \pi})$} is dimensionless, and the scale $M$ that takes care of the dimensions (see\,\eqref{deltaV}) turns out to be $M=\Lambda \equiv \frac{\sqrt{2 \pi}}{a}$, that is nothing but the Wilsonian UV physical cutoff of the theory (not an arbitrary mass scale). This observation will be useful for our following analysis.
 
Let us move now to the (Euclidean) action of scalar QED
\begin{align}\label{actionscalar}
	&S[\phi^*,\phi,A_\mu]=S_{\rm em}[A_\mu]+S_{0}[\phi^*,\phi]+S_{\rm int}[\phi^*,\phi,A_\mu]\nonumber\\
	&\equiv\frac 14\int \dd[4]{x}\, F_{\mu \nu}F^{\mu \nu}+\int \dd[4]{x} \left[\partial_\mu\phi^* \partial^\mu\phi+m^2\phi^*\phi\right]+\int \dd[4]{x}\left[ j_\mu A^\mu+e^2A_\mu A^\mu \phi^*\phi\right]\,,
\end{align}
where $F_{\mu \nu}=\partial_\mu A_\nu-\partial_\nu A_\mu$ and $j_\mu=ie(\phi\,\partial_\mu\phi^*-\phi^*\partial_\mu\phi)$. The Euler-Heisenberg correction $\delta S_{_{\rm EH}}^\phi$ to the free Maxwell action $S_{\rm em}$ is obtained integrating over the scalar fields (see \eqref{dimless})
\begin{equation}\label{EHactionsc}
	\delta S_{_{\rm EH}}^\phi=-\log\int  \text{\small$\mathcal D$}\text{\footnotesize$(\phi^*/\Lambda)$}  \text{\small$\mathcal D$}\text{\footnotesize$(\phi/\Lambda)$} \, e^{-(S_{0}[\phi^*,\phi]+S_{\rm int}[\phi^*,\phi,A_\mu])}\,.
\end{equation}
Performing the Gaussian integrations over $\phi$ and $\phi^*$ we get 
\begin{equation}\label{EHaction2}
	\delta S_{_{\rm EH}}^\phi=\log\det\Big(\frac{-D_\mu D^\mu+m^2}{\Lambda^2}\Big)=\Tr\log\Big(\frac{-D_\mu D^\mu+m^2}{\Lambda^2}\Big)\,,
\end{equation}
where $D_\mu=\partial_\mu + i e A_\mu$ is the covariant derivative.

As anticipated in the Introduction, we now identify the Wilsonian UV physical cutoff $\Lambda$ with a hard cut over the eigenvalues $\lambda_n^\phi$ of the operator $-D_\mu D^\mu+m^2$ that are included in the calculation of the trace in\,\eqref{EHaction2}, i.e.\,\,we retain only the eigenvalues such that $\lambda_n^\phi\leq\Lambda^2$. From a physical point of view, we are considering quantum fluctuations up to the maximal energy scale $\Lambda$ (down to the minimal distance $a$). Indicating with ${\rm D}_n^\phi$ the degeneracy of $\lambda_n^\phi$, we have 
\begin{align}\label{deltaSscalar2}
	\delta S_{_{\rm EH}}^\phi=\sum_{\lambda_n^\phi \leq \Lambda^2} {\rm D}_n^\phi\log \frac{\lambda_n^\phi}{\Lambda^2}\,.
\end{align}

In the next section, we specialize to the case of a constant magnetic (electric) field background. After solving the corresponding eigenvalue problem for the operator $-D_\mu D^\mu+m^2$, we calculate $\delta S_{_{\rm EH}}^\phi$ in\,\eqref{deltaSscalar2} for this background.

\subsection{Eigenvalue problem and calculation of $\delta S_{_{\rm EH}}^{\phi}$}
\label{eig-scalar}

In the present section, we consider the case of a constant magnetic field background $B\neq0$ (vanishing electric field, $E=0$). As it is clear from the explicit expression of the eigenvalues and of the corresponding degeneracies of $-D_\mu D^\mu+m^2$ (see\,\eqref{eigvalues3} and\,\eqref{degscalar2}), in Euclidean signature the result for $\delta S_{_{\rm EH}}^{\phi}$ in the case of a constant electric field is obtained through the replacement $B \to E$. Differences appear after analytic continuation to Lorentzian signature.

Taking now the background gauge field $A_\mu$ to be
\be \label{AmuE}
A_0=A_2=A_3=0 \, , \quad A_1= -B x_2 \, ,
\ee
the eigenvalue problem for the operator $-D_\mu D^\mu+m^2$ reads
\begin{equation}\label{eigenprobE}
	(-\partial_0^2-\partial_2^2-\partial_3^2-(\partial_1-ieB x_2)^2+m^2)\phi(x)=\lambda\phi(x)\,.
\end{equation}
This coincides with the eigenvalue problem for a free (non-relativistic) particle of mass $1/2$ in the $(x_0,x_3)$ plane, and with the Landau-level problem with frequency $2e B$ in the $(x_1,x_2)$ plane. The eigenvalues are
\begin{equation}\label{eigvalues3}
	\lambda_{n,k_0,k_3}^\phi=k_0^2+k_3^2+2e B \Big(n + \frac{1}{2}\Big)+m^2\,,\quad n=0,1,\dots\,;\,\,\,k_i=\frac{2\pi n_i}{L}\,,\,n_i\in\mathbb{Z}\,,
\end{equation}
where $L$ is the size of the quantization box (in the following we treat $k_i$ as continuum variables as usual). From now on we use the notation $k^2\equiv k_0^2+k_3^2$. The  eigenfunctions corresponding to the eigenvalues \eqref{eigvalues3} are
\begin{equation}\label{eigscalar}
	\phi_{n,k_0,k_3;p_1}=\text{\small$\bigg( \frac{2^{n}\, n!}{\sqrt{\pi}} \bigg)^{-\frac{1}{2}}$} \,\sqrt{e B\,}\, \text{\small$H_{n} \bigg(\sqrt{e B\,} \bigg(x_2 - \frac{p_1}{e B} \bigg)\bigg)e^{-\frac{e B}{2} \big(x_2 -\frac{p_1}{e B}\big)^2}$}\,e^{i p_1 x_1} e^{i (k_0 x_0+ k_3 x_3)}\,,
\end{equation}
where $H_n(x)$ are Hermite functions and $p_1\equiv\frac{2\pi n_1}{L}$, with $n_1$ integer. It is important to stress that $n_1$ is bounded. Actually, $\frac{p_1}{e B}$ is the center of the Gaussian that appears in $\phi_{n,k_0,k_3;p_1}$, and lies in the range $[-L/2,L/2]$ of the coordinate $x_2$, namely
\begin{equation}\label{p2}
	p_1^{\rm min}\equiv-\frac{e B}{2}L\leq p_1\leq\frac{e B}{2}L\equiv p_1^{\rm max}\,.
\end{equation}
Therefore, the number of linearly independent eigenfunctions associated to the eigenvalue\, $\lambda_{n,k_0,k_3}^\phi$\, is
\begin{equation}\label{degscalar}
	\frac{L}{2\pi}(p_1^{\rm max}-p_1^{\rm min})=\frac{e B}{2\pi} L^2\,,
\end{equation}
i.e. each eigenvalue $\lambda_{n,k_0,k_3}^\phi$ has degeneracy
\begin{equation}\label{degscalar2}
	{\rm D}_{n,k_0,k_3}^\phi=\frac{e B}{2\pi} L^2\,.
\end{equation}

Having found the eigenvalues $\lambda_{n,k_0,k_3}^\phi$ and the corresponding degeneracies ${\rm D}_{n,k_0,k_3}^\phi$, we can now calculate the Euler-Heisenberg correction $\delta S_{_{\rm EH}}^\phi$ to the free Maxwell action. As said above, we implement the physical cutoff $\Lambda$ as a hard cut on the spectrum, $\lambda_{n,k_0,k_3}^\phi \leq \Lambda^2$ (see\,\eqref{deltaSscalar2}). For the corresponding Lagrangian $\delta\mathcal L_{_{\rm EH}}^\phi=\delta S_{_{\rm EH}}^\phi/V$ (with $V=L^4$),  we have
\begin{align}\label{dSscalar2}
	\delta\mathcal L_{_{\rm EH}}^\phi=& \frac{e B}{2\pi} \sum_{n=0}^N\int^{(\text{K}_n)}\frac{\dd^2k}{(2\pi)^2} \log\Big(\frac{k^2+e B (2n + 1)+m^2}{\Lambda^2}\Big)\,,
\end{align}
where $N \equiv \lfloor \frac{\Lambda^2-m^2}{2 e B} - \frac12 \rfloor$ (with $\lfloor x\rfloor$ the floor of $x$), $\text{K}_n\equiv\sqrt{\Lambda^2-m^2 - eB(2n+1)}$ \,, and $\int^{(\text{K}_n)}$ indicates that the integration is over the circle of radius $\text{K}_n$.

Performing the integration over $k$ and the sum over $n$ we get
\begin{align}
	\delta\mathcal L_{_{\rm EH}}^\phi&=-\frac{e B}{8 \pi^2} (N+1) \Big(\Lambda ^2-m^2 \log \frac{\Lambda ^2}{2 e B}\Big)+\frac{e^2 B^2}{4 \pi^2} \Big[ \frac{N+1}{2} (N-\log (2 \pi )) \log \frac{\Lambda ^2}{2 e B} \nonumber \\
	&+\psi ^{(-2)}\Big(\frac{m^2}{2 e B}+\frac 12\Big) - \psi ^{(-2)}\Big(\frac{m^2}{2 e B}+\frac 32 + N\Big)\Big]\,,
\end{align}
where $\psi ^{(-n)}(z)$ (with $n$ positive integer) is the polygamma function of negative order $-n$ defined as\,\cite{Adamchik} ($\Gamma(z)$ is the Euler Gamma function)
\begin{equation}
	\psi ^{(-n)}\left(z\right)=\frac{1}{(n-2)!}\int_{0}^z\dd{t}(z-t)^{n-2} \log\,\Gamma\left(t\right)\qquad\qquad \text{for}\,\,\Re(z)>0\,.
\end{equation}
Defining
\begin{equation}\label{delta}
\xi \equiv  \Big( \frac{\Lambda^2-m^2}{2 e B} - \frac12 \Big) - \Big\lfloor \frac{\Lambda^2-m^2}{2 e B} - \frac12 \Big\rfloor\,,
\end{equation}
$N$ can be written as
\begin{equation}
    N=\frac{\Lambda^2-m^2}{2 e B} - \frac12 - \xi\,.
\end{equation}
We now expand $\delta \mathcal L_{_{\rm EH}}^\phi$ for $m^2, e B\ll\Lambda^2$. We get (note that according to\,\eqref{delta}, $0\leq\xi<1$ independently of the specific values of $\Lambda$, $e B$ and $m^2$)
\begin{align}\label{scalarresult}
	\delta\mathcal L_{_{\rm EH}}^\phi&=-\frac{\Lambda ^4}{64 \pi ^2}+\frac{m^2\Lambda ^2}{16 \pi ^2}-\frac{m^4}{32\pi^2}\Big(\log\frac{\Lambda^2}{m^2}+\frac32\Big)\nn\\
	&+\frac{m^4}{32\pi^2}\Big(\frac32-\log\frac{m^2}{2eB}\Big)-\frac{m^2 e B}{16\pi^2}\log(2\pi)\nn\\
	&+\frac{e^2 B^2}{96\pi^2}\Big[\log\frac{\Lambda^2}{m^2}-6\log(2\pi A^4)\Big]+\frac{e^2 B^2}{96\pi^2}\Big[\log\frac{m^2}{2eB}+24\psi^{(-2)}\Big(\frac{m^2}{2eB}+\frac12\Big)\Big]\nn\\
	&+\frac{e^2 B^2}{24\pi^2}\Big[\frac{e B}{\Lambda^2}\xi(1-\xi)(1-2\xi)+\frac{e^2 B^2}{\Lambda^4}\Big(\xi^2(1-\xi)^2-\frac{1}{30}\Big)+\mathcal O\Big(\frac{e^3 B^3}{\Lambda^6}\Big)\Big]\,.
\end{align}
The above equation is our result for the Euler-Heisenberg correction $\delta S_{_{\rm EH}}^\phi$ due to a scalar field (see\,\eqref{actionscalar}). It is one of the central results of the present work. Before delving into a detailed discussion of the different terms in\,\eqref{scalarresult}, we now make some comments. 

For the first time the calculation of $\delta S_{_{\rm EH}}$ has been performed resorting to a {\it hard} cutoff $\Lambda$, which has a precise physical meaning. Within the Wilsonian EFT paradigm, $\Lambda$ is the ultimate scale above which the theory has to be replaced by its UV completion. 
Moreover, it is important to stress that, since $\Lambda$ is implemented as a cut over the eigenvalues of the fluctuation operator (that are gauge invariant), our result\,\eqref{scalarresult} is gauge invariant.

In the literature, the calculation of $\delta S_{_{\rm EH}}$ has been performed in many other ways \cite{Heisenberg:1936nmg, Schwinger:1951nm,Dittrich:1975au, Dunne:2004nc, Dunne:2012vv, Weisskopf:1936hya, Gies:2016yaa}, resorting to regularization schemes that range from Pauli-Villars regularization to zeta function regularization, proper-time method, etc. As it is the case for the Wilsonian UV physical cutoff $\Lambda$ introduced above, these techniques preserve gauge invariance. However, contrary to the use of the spectral hard cutoff, which has the precise physical meaning stressed above, these regularization techniques do not have an immediate physical interpretation. 

In this respect, it is worth to mention that Pauli and Villars themselves described their regularization method as a \vv formalistic'' \vv transitional stage of the theory'', awaiting a truly physical way to preserve gauge invariance at the quantum level \cite{Pauli:1949zm}. Similarly, in his seminal paper \cite{Schwinger:1951nm} Schwinger refers to the proper-time method as a technique to preserve gauge and Lorentz invariance, without delving into its physical interpretation\footnote{\label{ptfootnote} Apart from these historical remarks, it is worth to observe that from the expression of the functional determinant of a given operator $\mathcal O$ regularized in terms of a proper-time cutoff $1/\Lambda^2$ ($\lambda_n$ eigenvalues of $\mathcal O$ and D$_n$ corresponding degeneracies), $	\log\det(\mathcal O)=-\sum_{n}^{\infty}{\rm D}_n\int_{1/\Lambda^2}^{\infty}\frac{\dd s}{s}e^{-s\lambda_n}=-\sum_{n}^{\infty}{\rm D}_n\int_{0}^{\infty}\frac{\dd s}{s}\,\text{\scriptsize$\Theta\big(s-1/\Lambda^2\big)$}e^{-s\lambda_n}$ \cite{Schwinger:1951nm}\,,
it is seen that $1/\Lambda^2$ implements a smooth suppression of the eigenvalues of the operator $\mathcal O$ greater than $\Lambda^2$. Similar considerations hold for smooth versions of the proper-time cutoff where $\text{\scriptsize$\Theta\big(s-1/\Lambda^2\big)$}$ is replaced by a smooth function $\rho(s,\Lambda)$ \cite{Oleszczuk:1994st, Liao:1994fp}.}. 

Let us go back to\,\eqref{scalarresult}. The terms in the first line are just inessential vacuum energy contributions. We consider now those in the second and third line. As it can be easily verified, these terms (that we obtained resorting to the physical {\it hard} cutoff $\Lambda$) coincide with the result for $\delta S_{_{\rm EH}}^\phi$ that is obtained resorting to proper-time regularization (see for instance \cite{Dittrich:1975au, Dunne:2004nc,Dunne:2012vv} where the result is written in Lorentzian signature. Here we are using Euclidean signature, but in the case of a magnetic field background the difference consists simply in an overall minus sign),
\begin{align}\label{ptres}
	\big[\delta\mathcal L_{_{\rm EH}}^\phi\big]_{_{\rm pt}}=-\frac{1}{16\pi^2}\int_{1/\Lambda^2}^{+\infty}\frac{\dd s}{s^3}\Big(\frac{e B s}{\sinh(e B s)}-1\Big)e^{-sm^2}\,.
\end{align}

We want to consider now the case of a constant electric field background $E$. As said at the beginning of the present section, the (Euclidean) result for this case ($E\neq0$, $B=0$) is obtained by making in\,\eqref{scalarresult} the replacement $B\to E$. Considering this case, and performing the analytic continuation of Eq.\,\eqref{scalarresult} to Lorentzian signature ($E\to -i E$ and $\delta\mathcal L_{_{\rm EH}}^\phi\to-\delta\mathcal L_{_{\rm EH}}^\phi$), it is well-known that $\delta\mathcal L_{_{\rm EH}}^\phi$ develops an imaginary part related to the production of $e^+ e^-$ pairs. Sticking as before only to the second and third line of this equation, since (as stressed above) in Euclidean signature these two lines coincide with\,\eqref{ptres}, we have that (after analytic continuation) the imaginary part of our $\delta\mathcal L_{_{\rm EH}}^\phi$ coincides with the imaginary part of $\big[\delta\mathcal L_{_{\rm EH}}^\phi\big]_{_{\rm pt}}$ (see for instance \cite{Schwinger:1951nm, Dittrich:1975au, Dunne:2004nc, Dunne:2012vv}), apart from contributions from the fourth line of\,\eqref{scalarresult} that are generically cutoff-suppressed.

Let us finally comment on the terms in this fourth line. Since they depend on $\xi$, which is periodic in $1/e B$ with period $2(\Lambda^2-m^2)^{-1}$ (see\,\eqref{delta}), they are periodic in $1/e B$ with the same period. Moreover, since $\xi$ lies in the range $0\leq\xi<1$, for values of the magnetic field $B$ such that $e B\ll\Lambda^2$, these terms are cutoff suppressed and our result\,\eqref{scalarresult} for $\delta\mathcal L_{_{\rm EH}}^\phi$ coincides with the result $\big[\delta\mathcal L_{_{\rm EH}}^\phi\big]_{_{\rm pt}}$ obtained resorting to proper-time regularization.

However, we have to stress again that the UV physical cutoff $\Lambda$ is not a technical device introduced to render finite ill-defined expressions involved in the calculation of quantum fluctuations. It has instead the deep physical meaning of being the ultimate scale of the theory, above which the latter has to be replaced by its UV completion. Therefore, going back to our result\,\eqref{scalarresult}, we see that for strong magnetic fields, more precisely for\,values of $e B$ not much smaller than $\Lambda^2$, the terms in the fourth line of this equation (absent in usual calculations of $\delta\mathcal L_{_{\rm EH}}^\phi$, in particular in $\big[\delta\mathcal L_{_{\rm EH}}^\phi\big]_{_{\rm pt}}$) may become relevant and provide significant modifications to physical quantities. This point deserves further investigation both from a theoretical and phenomenological point of view (further comments are in section \ref{Conclusions}). 

In the next section, we perform a similar analysis for the case of fermionic QED.

\section{Euler-Heisenberg action. Fermionic QED}
\label{EH-fermionic}

Let us consider the (Euclidean) action for fermionic QED
\begin{align}\label{actionfermion}
	S[A,\psi,\bar \psi]&=S_{\rm em}[A_\mu]+S_{0}[\bar\psi,\psi]+S_{\rm int}[\bar\psi,\psi,A_\mu]\nonumber\\
	&\equiv\frac 14\int \dd[4]{x}\, F_{\mu \nu}F^{\mu \nu}+\int \dd[4]{x}\bar \psi \left(\slashed \partial+m\right)\psi+i e\int \dd[4]{x}\bar \psi \gamma_\mu\psi A^\mu\,,
\end{align}
where $\slashed{\partial}=\gamma^\mu \partial_\mu$. $\gamma^\mu$ are the (Euclidean) gamma matrices,
\begin{align}\label{gammamatrices}
	\gamma^0=
	\begin{pmatrix}
		0 & \mathbb 1_{2} \\
		\mathbb 1_{2} & 0
	\end{pmatrix}
	\qquad 
	\gamma^i=
	\begin{pmatrix}
		0 &	-i\sigma^i \\
		i\sigma^i & 0
	\end{pmatrix}\,,
\end{align}
and
\begin{equation}\label{gamma5}
	\gamma_5=\gamma^0\gamma^1\gamma^2\gamma^3=
	\begin{pmatrix}
		-\mathbb 1_{2} &	0 \\
		0 & \mathbb 1_{2}
	\end{pmatrix}\,.
\end{equation}

Similarly to the scalar QED case considered in the previous section, the Euler-Heisenberg action $\delta S_{_{\rm EH}}^\psi$ is ($\slashed D=\gamma^\mu D_\mu$)
\begin{equation}\label{EHactionferm2}
	\delta S_{_{\rm EH}}^\psi=-\log\det\Big(\frac{\slashed D+m}{\Lambda}\Big)=-\Tr\log\Big(\frac{\slashed D+m}{\Lambda}\Big)\,.
\end{equation}
Again we implement the Wilsonian UV physical cutoff \,$\Lambda$\, with a hard cut over the eigenvalues $\lambda_n^\psi$ of the operator $\slashed D+m$. Being $\slashed D$ anti-hermitian, its eigenvalues $\lambda_n^\psi-m$ are purely imaginary. For each of them, the spectrum of $\slashed D$ also contains the complex conjugate eigenvalue $(\lambda_n^\psi)^*-m$. The spectral cut is then implemented on $|\lambda_n^\psi|^2$ as $|\lambda_n^\psi|^2\leq\Lambda^2$. We have then (${\rm D}_n^\psi$ is the degeneracy of $\lambda_n^\psi$)
\begin{align}\label{deltaferm2}
	\delta S_{_{\rm EH}}^\psi=-\sum_{|\lambda_n^\psi|^2 \leq \Lambda^2} {\rm D}_n^\psi\log \frac{|\lambda_n^\psi|^2}{\Lambda^2}\,.
\end{align}

In the next section, we solve the eigenvalue problem for the operator $\slashed D+m$ in the case of a constant magnetic background, and calculate the corresponding $\delta S_{_{\rm EH}}^\psi$ in\,\eqref{deltaferm2}.

\subsection{Eigenvalue problem and calculation of $\delta S_{_{\rm EH}}^{\psi}$}

Similarly to what we did for scalar QED, we now consider the case of a constant magnetic field background $B\neq0$ (in Euclidean signature, the result for $\delta S_{_{\rm EH}}^\psi$ for a constant electric field background is obtained through the replacement $B\to E$). 

We now perform the calculation taking for the gauge field $A_\mu$
\be \label{AmuEferm}
A_0=A_2=A_3=0 \, , \qquad A_1=-B x_2  \,.
\ee
The eigenvalue problem for the Dirac operator $\slashed D+m$ then reads
\begin{equation}\label{eigenprobEferm}
	(\gamma^\mu\partial_\mu-ieB\gamma^1x_2+m)\psi(x)=\lambda\psi(x)\,.
\end{equation}
As for the case of scalar QED of section \ref{EH-scalar}, this problem can be solved analytically. The details of the calculation are given in Appendix\,\ref{app1}. The eigenvalues are
\begin{equation}\label{eigvalues3ferm}
	\lambda^\psi_{n,k_0,k_3}=m\pm i\sqrt{k_0^2+k_3^2+2 e B n}\,,\quad n=0,1,\dots\,;\,\,\,k_i=\frac{2\pi n_i}{L}\,,\,n_i\in\mathbb{Z}\,.
\end{equation}
From now on we use the notation $k^2 \equiv k_0^2+k_3^2$. To obtain the  eigenfunctions corresponding to the eigenvalues \eqref{eigvalues3ferm}, we treat separately the cases $n=0$ and $n\neq0$. Let us begin from the latter.

\vskip 6pt

\noindent
{\it Case $n\neq0$} 

\vskip 3pt

Let us begin by noting that, for any given $n, k_0, k_3$, we have two complex conjugate eigenvalues of $\slashed D+m$, namely $\lambda^\psi_{n,k_0,k_3}=m+ i\sqrt{k^2+2 e B n}$ and $(\lambda^\psi_{n,k_0,k_3})^*$ (see\,\eqref{eigvalues3ferm}). As shown in Appendix \ref{app1}, the eigenfunctions associated to $(\lambda^\psi_{n,k_0,k_3})^*$ are obtained applying the matrix $\gamma_5$ to the eigenfunctions associated to $\lambda^\psi_{n,k_0,k_3}$.  A set of linearly independent eigenfunctions corresponding to the eigenvalue $\lambda^\psi_{n,k_0,k_3}$ is given by (see\,\eqref{eigphi1} and\,\eqref{eigphi2})
\begin{align}
	\big(\psi^{\text{\tiny$(1)$}}_{\lambda}\big)_{n,k_0,k_3;p_1}&=\begin{pmatrix}
		\phi_{n-1,k_0,k_3;p_1} \\
		0\\
		\frac{k_0+ik_3}{\sqrt{k^2+2e Bn}}\phi_{n-1,k_0,k_3;p_1}\\
		\frac{-i\sqrt{2e B n\,}}{\sqrt{k^2+2eBn}}\,\phi_{n,k_0,k_3;p_1}
	\end{pmatrix}\label{eigphi1testo}\\	\big(\psi^{\text{\tiny$(2)$}}_{\lambda}\big)_{n,k_0,k_3;p_1}&=\begin{pmatrix}
		0\\
		\phi_{n,k_0,k_3;p_1}\\
		\frac{-i\sqrt{2e B n\,}}{\sqrt{k^2+2eBn}}\,\phi_{n-1,k_0,k_3;p_1}\\
		\frac{k_0-ik_3}{\sqrt{k^2+2eBn}}\phi_{n,k_0,k_3;p_1}
	\end{pmatrix}\label{eigphi2testo}\,.
\end{align}
We have already encountered the functions $\phi_{n,k_0,k_3;p_1}$ when studying the scalar case, see Eq.\,\eqref{eigscalar} and comments below. In particular, $p_1$ is the degeneracy index and lies in the range $p_1^{\rm min}\equiv-\frac{e B}{2}L\leq p_1\leq\frac{e B}{2}L\equiv p_1^{\rm max}$ (see\,\eqref{p2} and comments below\,\eqref{solphi2}). Therefore, the number of linearly independent eigenfunctions {\small $\big(\psi^{\text{\tiny$(1)$}}_{\lambda}\big)_{n,k_0,k_3;p_1}$ $\big($$\big(\psi^{\text{\tiny$(2)$}}_{\lambda}\big)_{n,k_0,k_3;p_1}$$\big)$} associated to the eigenvalue\, $\lambda_{n,k_0,k_3}^\psi$\, is
\begin{equation}\label{deg4}
	\frac{L}{2\pi}(p_1^{\rm max}-p_1^{\rm min})=\frac{e B}{2\pi} L^2\,,
\end{equation}
and the degeneracy of the eigenvalue $\lambda_{n,k_0,k_3}^\psi$ is
\begin{equation}\label{deg5}
	D_{n,k_0,k_3}^\psi=\frac{e B}{\pi} L^2\,.
\end{equation}
Finally, the eigenfunctions $\big(\psi^{\text{\tiny$(1)$}}_{\lambda^*}\big)_{n,k_0,k_3;p_1}$ and\, $\big(\psi^{\text{\tiny$(2)$}}_{\lambda^*}\big)_{n,k_0,k_3;p_1}$ associated to the eigenvalue $(\lambda_{n,k_0,k_3}^\psi)^*$ are (as anticipated) $\big(\psi^{\text{\tiny$(1)$}}_{\lambda^*}\big)_{n,k_0,k_3;p_1}=-\gamma_5 \big(\psi^{\text{\tiny$(1)$}}_{\lambda}\big)_{n,k_0,k_3;p_1}$ and $\big(\psi^{\text{\tiny$(2)$}}_{\lambda^*}\big)_{n,k_0,k_3;p_1}=-\gamma_5 \big(\psi^{\text{\tiny$(2)$}}_{\lambda}\big)_{n,k_0,k_3;p_1}$, and the eigenvalue $(\lambda_{n,k_0,k_3}^\psi)^*$ has the same degeneracy\,\eqref{deg5} as $\lambda_{n,k_0,k_3}^\psi$.

\vskip 8pt

\noindent
{\it Case $n=0$} 

\vskip 3pt

Let us consider now the eigenvalues $\lambda_{0,k_0,k_3}^\psi=m+ i k$. A set of linearly independent eigenfunctions corresponding to $\lambda_{0,k_0,k_3}^\psi$ and $(\lambda_{0,k_0,k_3}^\psi)^*$ is (see\,\eqref{eigphiq1} and\,\eqref{eigphiq2})
\begin{align}
		\big(\psi_{\lambda}\big)_{0,k_0,k_3;p_1}=\begin{pmatrix}
			0\\
			\phi_{0,k_0,k_3;p_1}\\
			0\\
			\frac{k_0-ik_3}{k}\,\phi_{0,k_0,k_3;p_1}
		\end{pmatrix}\label{eigphi0E2}
\end{align}
and
\begin{align}
		\big(\psi^{\text{\tiny$(2)$}}_{\lambda^*}\big)_{0,k_0,k_3;p_1}=\begin{pmatrix}
			0\\
			\phi_{0,k_0,k_3;p_1}\\
			0\\
			-\frac{k_0-ik_3}{k}\,\phi_{0,k_0,k_3;p_1}\end{pmatrix}\,,\label{eigphi0E}
\end{align}

\noindent
respectively. The degeneracy of both $\lambda_{0,k_0,k_3}^\psi$ and $(\lambda_{0,k_0,k_3}^\psi)^*$ is $D_{0,k_0,k_3}^\psi=\frac{e B}{2\pi} L^2$ (see comments below\,\eqref{eigphiq2}).

\vskip 3pt

The complete set of orthogonal eigenfunctions that spans the field $\psi$ space is then (see Eqs.\,\eqref{eigphi1testo}, \eqref{eigphi2testo}, \eqref{eigphi0E2} and\,\eqref{eigphi0E})
\begin{equation}\label{basistesto}
	\big(\psi^{\text{\tiny$(1)$}}_{\lambda}\big)_{n,k_0,k_3;p_1},\big(\psi^{\text{\tiny$(1)$}}_{\lambda^*}\big)_{n,k_0,k_3;p_1},\big(\psi^{\text{\tiny$(2)$}}_{\lambda}\big)_{n,k_0,k_3;p_1},\big(\psi^{\text{\tiny$(2)$}}_{\lambda^*}\big)_{n,k_0,k_3;p_1},\big(\psi^{\text{\tiny$(2)$}}_{\lambda}\big)_{0,k_0,k_3;p_1},\big(\psi^{\text{\tiny$(2)$}}_{\lambda^*}\big)_{0,k_0,k_3;p_1}\,.
\end{equation}
Similarly, the $\bar\psi$ space is spanned by
\begin{equation}\label{basis2testo}
	\big(\psi^{\text{\tiny$(1)$}}_{\lambda}\big)^\dagger_{n,k_0,k_3;p_1},\big(\psi^{\text{\tiny$(1)$}}_{\lambda^*}\big)^\dagger_{n,k_0,k_3;p_1},\big(\psi^{\text{\tiny$(2)$}}_{\lambda}\big)^\dagger_{n,k_0,k_3;p_1},\big(\psi^{\text{\tiny$(2)$}}_{\lambda^*}\big)^\dagger_{n,k_0,k_3;p_1},\big(\psi^{\text{\tiny$(2)$}}_{\lambda}\big)^\dagger_{0,k_0,k_3;p_1},\big(\psi^{\text{\tiny$(2)$}}_{\lambda^*}\big)^\dagger_{0,k_0,k_3;p_1}\,.
\end{equation}

For the Euler-Heisenberg Lagrangian $\delta \mathcal L_{_{\rm EH}}^\psi=\delta S_{_{\rm EH}}^\psi/V$ we then have (see\,\eqref{deltaferm2})
\begin{align}\label{dSferm2}
	\delta \mathcal L_{_{\rm EH}}^\psi=&-\frac{e B}{\pi}\sum_{n=1}^{N}\int^{(\text{K}_n)}\frac{\dd^2k}{(2\pi)^2} \log\Big(\frac{k^2+2e Bn+m^2}{\Lambda^2}\Big)-\frac{e B}{2\pi}\int^{(\Lambda)}\frac{\dd^2k}{(2\pi)^2}\log\Big(\frac{k^2+m^2}{\Lambda^2}\Big)\,,
\end{align}
where $N \equiv \lfloor \frac{\Lambda^2-m^2}{2 e B} \rfloor$, $\text{K}_n\equiv\sqrt{\Lambda^2-m^2 - 2eBn}$ \,, and $\int^{(\text{K}_n)}$ ($\int^{(\Lambda)}$) indicates that the integration is over the circle of radius $\text{K}_n$ (radius $\Lambda$).
Performing the integration over $k$ and the sum over $n$ we get
{\small\begin{align}
	\delta\mathcal L_{_{\rm EH}}^\psi&=\frac{e B}{8 \pi^2} \Big[(2N+1)\Lambda^2-m^2\Big(2N\log\frac{\Lambda^2}{2 e B}+\log\frac{\Lambda^2}{m^2}+1\Big)\Big]\nonumber\\	
	&-\frac{e^2 B^2}{2 \pi^2} \Big[ \frac{N}{2}\Big( (N+1) \log \frac{\Lambda ^2}{2 e B}+\log(2\pi)\Big) +\psi ^{(-2)}\Big(\frac{m^2}{2 e B}+1\Big) - \psi ^{(-2)}\Big(\frac{m^2}{2 e B}+1 + N\Big)\Big]\,.
\end{align}}

Similarly to the previous section, we now define
\begin{equation}\label{deltaferm}
	\xi \equiv  \Big( \frac{\Lambda^2-m^2}{2 e B} \Big) - \Big\lfloor \frac{\Lambda^2-m^2}{2 e B} \Big\rfloor\,,
\end{equation}
so that $N$ can be written as
\begin{equation}
	N=\frac{\Lambda^2-m^2}{2 e B} - \xi\,.
\end{equation}
Expanding $\delta \mathcal L_{_{\rm EH}}^\psi$ for $m^2, e B\ll\Lambda^2$ (noting that $0\leq\xi<1$), we find
\begin{align}\label{fermionresult}
	\delta\mathcal L_{_{\rm EH}}^\psi&=\frac{\Lambda ^4}{32 \pi ^2}-\frac{m^2\Lambda ^2}{8 \pi ^2}+\frac{m^4}{16\pi^2}\Big(\log\frac{\Lambda^2}{m^2}+\frac32\Big)\nn\\
	&+\frac{m^4}{16\pi^2}\Big(\log\frac{m^2}{2eB}-\frac32\Big)+\frac{m^2 e B}{8\pi^2}\Big(\log\frac{m^2}{2eB}+\log(2\pi)-1\Big)\nn\\
	&+\frac{e^2 B^2}{24\pi^2}\Big[\log\frac{\Lambda^2}{m^2}+6\log(2\pi A^2)\Big]+\frac{e^2 B^2}{24\pi^2}\Big[\log\frac{m^2}{2eB}-12\psi^{(-2)}\Big(\frac{m^2}{2eB}+1\Big)\Big]\nn\\
	&-\frac{e^2 B^2}{12\pi^2}\Big[\frac{e B}{\Lambda^2}\xi(1-\xi)(1-2\xi)+\frac{e^2 B^2}{\Lambda^4}\Big(\xi^2(1-\xi)^2-\frac{1}{30}\Big)+\mathcal O\Big(\frac{e^3 B^3}{\Lambda^6}\Big)\Big]
\end{align}
The above equation is our result for the Euler-Heisenberg Lagrangian $\delta \mathcal L_{_{\rm EH}}^\psi$, another central outcome of the present work. Comparing\,\eqref{fermionresult} to our previous result\,\eqref{scalarresult} for $\delta \mathcal L_{_{\rm EH}}^\phi$ of scalar QED, it is clear that the comments made below\,\eqref{scalarresult} hold true also for $\delta \mathcal L_{_{\rm EH}}^\psi$. Here, we simply stress that the second and third lines of\,\eqref{fermionresult} coincide (after analytic continuation to Lorentzian signature) with the result for the Euler-Heisenberg Lagrangian obtained in the 
pioneering work \cite{Schwinger:1951nm} by Schwinger. As for the scalar QED case, the terms in the first line are vacuum energy terms, while those in the fourth line are generically cutoff-suppressed terms, periodic in $1/e B$ (see the fourth line of\,\eqref{scalarresult} and comments below).

In the next section, we go back to the question raised at the beginning of section \ref{EH-scalar}, namely the issue of the scale $M$ that takes care of the dimensions of the fluctuation operator in loop calculations.

\section{Discussion}
\label{discussion}

Starting from the original phase space path integral formulation of a QFT, in section \ref{EH-scalar} we argued that the scale $M$ that takes care of the dimensions of the fluctuation operator in loop calculations is the Wilsonian UV physical cutoff $\Lambda$ ($M=\Lambda$). In the present section, we discuss the consequences of choosing $M \neq \Lambda$. We will start from the case of scalar QED showing that, if to take care of the dimensions of the operator  $-D_\mu D^\mu+m^2$ in\,\eqref{EHaction2} a generic fixed scale $M$ is used rather than the Wilsonian physical scale $\Lambda$ (i.e.\,\,if in\,\eqref{EHaction2} we replace $\Lambda^2$  with $M^2$ inside the logarithm), unphysical terms appear in $\delta S_{_{\rm EH}}^\phi$. We will see that similar considerations hold true for $\delta S_{_{\rm EH}}^\psi$. Moreover, we will show that $M=\Lambda$ is crucial to frame the proper-time method within the Wilsonian paradigm. 

Let us begin then with scalar QED. As before, $\Lambda$ is identified as a cut over the eigenvalues $\lambda_n$ of $-D_\mu D^\mu+m^2$ ($\lambda_n\leq\Lambda^2$). If we take $M \neq \Lambda$, for $\delta S_{_{\rm EH}}^\phi$ we have (${\rm D}_n$ degeneracy of $\lambda_n$)
\begin{align}\label{deltaSscalar2mu}
	\delta S_{_{\rm EH}}^\phi=\sum_{\lambda_n^\phi \leq \Lambda^2} {\rm D}_n\log \frac{\lambda_n^\phi}{M^2}\,.
\end{align}

\noindent
Specializing\,\eqref{deltaSscalar2mu} to the case of a constant magnetic field background $B$, the Euler-Heisenberg Lagrangian $\delta\mathcal L_{_{\rm EH}}^\phi$ turns out to be (see\,\eqref{delta} for the definition of $\xi\text{\small$(B)$}$)
\begin{align}\label{scalarresultmu}
	&\delta\mathcal L_{_{\rm EH}}^\phi=-\frac{\Lambda ^4}{64 \pi ^2}\Big(1-2\log\frac{\Lambda^2}{M^2}\Big)+\frac{m^2\Lambda ^2}{16 \pi ^2}\Big(1-\log\frac{\Lambda^2}{M^2}\Big)-\frac{m^4}{32\pi^2}\Big(\log\frac{\Lambda^2}{m^2}+\frac32-\log\frac{\Lambda^2}{M^2}\Big)\nn\\
	&+\frac{m^4}{32\pi^2}\Big(\frac32-\log\frac{m^2}{2eB}\Big)-\frac{m^2 e B}{16\pi^2}\log(2\pi)\nn\\
	&+\frac{e^2 B^2}{96\pi^2}\Big[\log\frac{\Lambda^2}{m^2}-3\log\frac{\Lambda^2}{M^2}-6\log(2\pi A^4)\Big]+\frac{e^2 B^2}{96\pi^2}\Big[\log\frac{m^2}{2eB}+24\psi^{(-2)}\Big(\frac{m^2}{2eB}+\frac12\Big)\Big]\nn\\
	&+\frac{e^2 B^2}{24\pi^2}\Big[3\xi(1-\xi)\log\frac{\Lambda^2}{M^2}+\frac{e B}{\Lambda^2}\xi(1-\xi)(1-2\xi)+\frac{e^2 B^2}{\Lambda^4}\Big(\xi^2(1-\xi)^2-\frac{1}{30}\Big)+\mathcal O\Big(\frac{e^3 B^3}{\Lambda^6}\Big)\Big]\,.
\end{align}

The comparison between\,\eqref{scalarresultmu} and\,\eqref{scalarresult} shows that, apart from inessential additional vacuum energy terms (first line of\,\eqref{scalarresultmu}), the differences due to the use of a generic fixed scale $M$ (rather than $\Lambda$) reside in the term $-3\log\frac{\Lambda^2}{M^2}$ in the third line  of\,\eqref{scalarresultmu} and in the term 	\,$\text{\small$3\,\xi\text{\footnotesize$(B)$}(1-\xi\text{\footnotesize$(B)$})$}\log\frac{\Lambda^2}{M^2}$\, in the fourth line. From a simple look to\,\eqref{scalarresultmu}, we see that the former term has a dramatic impact on the beta function $\beta(e)$ for the running charge. In fact, for $M=\Lambda$ this term vanishes and the beta function $\beta(e)$ is solely due to the term $\log\frac{\Lambda^2}{m^2}$ (third line) that leads to $\beta(e)=\frac{e^3}{48\pi^2}$, the well-known one-loop result for scalar QED. On the contrary, for generic $M\neq\Lambda$, the additional term $-3\log\Lambda^2$ modifies the beta function leading to $\beta(e)=-\frac{e^3}{36\pi^2}$, which not only has a different coefficient in front of $e^3$, but even worse it is opposite in sign (Landau pole $\to$ asymptotic freedom). Needless to say, this is an absurd result that further supports $M=\Lambda$ (see the discussion at the beginning of section \ref{EH-scalar}). Let us also observe that the other additional term due to $M\neq \Lambda$ (first term in the fourth line of\,\eqref{scalarresultmu}) is a term periodic in $1/e B$, with period $2(\Lambda^2-m^2)^{-1}$, that significantly changes the field dependence of $\delta\mathcal L_{_{\rm EH}}^\phi$, giving rise to spurious large amplitude oscillations.

We now repeat the same calculation with $M \neq \Lambda$ in the case of fermionic QED (see\,\eqref{deltaferm2}),
\begin{align}\label{deltaferm2mu}
	\delta S_{_{\rm EH}}^\psi=
	-\sum_{|\lambda_n^\psi|^2 \leq \Lambda^2} {\rm D}_n\log \frac{|\lambda_n^\psi|^2}{M^2}\,.
\end{align}
In the case of a constant magnetic field background $B$ we get
\begin{align}\label{fermionresultmu}
	&\delta\mathcal L_{_{\rm EH}}^\psi=\frac{\Lambda ^4}{32 \pi ^2}\Big(1-2 \log \frac{\Lambda^2}{M^2}\Big)-\frac{m^2\Lambda ^2}{8 \pi ^2}\Big(1- \log \frac{\Lambda^2}{M^2}\Big)+\frac{m^4}{16\pi^2}\Big(\log\frac{\Lambda^2}{m^2}+\frac32 - \log \frac{\Lambda^2}{M^2}\Big)\nn\\
	&+\frac{m^4}{16\pi^2}\Big(\log\frac{m^2}{2eB}-\frac32\Big)+\frac{m^2 e B}{8\pi^2}\Big(\log\frac{m^2}{2eB}+\log(2\pi)-1\Big)\nn\\
	&+\frac{e^2 B^2}{24\pi^2}\Big[\log\frac{\Lambda^2}{m^2}+6\log(2\pi A^2)\Big]+\frac{e^2 B^2}{24\pi^2}\Big[\log\frac{m^2}{2eB}-12\psi^{(-2)}\Big(\frac{m^2}{2eB}+1\Big)\Big]\nn\\
	&-\frac{e^2 B^2}{12\pi^2}\Big[3\xi(1-\xi)\log\frac{\Lambda^2}{M^2}+\frac{e B}{\Lambda^2}\xi(1-\xi)(1-2\xi)+\frac{e^2 B^2}{\Lambda^4}\Big(\xi^2(1-\xi)^2-\frac{1}{30}\Big)+\mathcal O\Big(\frac{e^3 B^3}{\Lambda^6}\Big)\Big]\,.
\end{align}
 
Differently from what happens for scalar QED, in this case there are no additional terms of the kind $e^2 B^2 \log \Lambda^2$ that modify the one-loop beta function $\beta(e)$. This is due to the difference between\, $\lambda_n^\phi=k^2+e B (2n + 1)+m^2$\, in \eqref{deltaSscalar2mu}  ($\lambda_n^\phi$ eigenvalues of $-D_\mu D^\mu+m^2$) and\, $|\lambda_n^\psi|^2=k^2+2e B n +m^2$\, in \eqref{deltaferm2mu} ($\lambda_n^\psi$ and $(\lambda_n^\psi)^*$ eigenvalues of $\slashed D+m$). The eigenvalue $\lambda_n^\phi$ has an extra $eB$ term compared to $|\lambda_n^\psi|^2$ that, as it can be easily seen, is responsible for the term \,$-3\log \Lambda^2$\, that appears in $\delta\mathcal L_{_{\rm EH}}^\phi$. 
However, as for the scalar QED case, the choice $M\neq \Lambda$ gives rise to a term periodic in $1/e B$ with period $2(\Lambda^2-m^2)^{-1}$ (first term in the fourth line of\,\eqref{fermionresultmu}), that again sensibly changes the field dependence of $\delta\mathcal L_{_{\rm EH}}^\psi$ giving rise to spurious large amplitude oscillations.

Let us consider now the issue of the scale $M$ to be used in the fluctuation determinant from a different perspective, comparing our calculation in terms of a hard cutoff $\Lambda$ to its definition within the proper-time method. We show that, taking $M=\Lambda$, the proper-time definition is recovered from the expression written in terms of the Wilsonian UV hard cutoff $\Lambda$. Conversely, we will see that for a generic $M \neq \Lambda$ this connection cannot be established.

We focus on $\delta S_{_{\rm EH}}^\phi$ in \eqref{deltaSscalar2mu} (the same analysis can be repeated for $\delta S_{_{\rm EH}}^\psi$) and write $\log \frac{\lambda_n^\phi}{M^2}$ using the identity\, $\log \frac{A}{B} = \int_0^\infty \frac{\dd s}{s} (e^{-s B}-e^{-s A})$,
\begin{equation}\label{ptidentity1}
\delta S_{_{\rm EH}}^\phi=\sum_{\lambda_n^\phi\leq\Lambda^2}{\rm D}_n\log\frac{\lambda_n^\phi}{M^2}=\sum_{\lambda_n^\phi\leq\Lambda^2}{\rm D}_n\int_0^{\infty}\frac{\dd s}{s}(e^{-sM^2}-e^{-s\lambda_n^\phi})\,.
\end{equation}
The latter can be trivially rewritten as
\begin{equation}\label{ptidentity2}
	\delta S_{_{\rm EH}}^\phi=	\lim_{\Lambda_{\rm pt}\to\infty}\int_{1/\Lambda_{\rm pt}^2}^{\infty}\frac{\dd s}{s}\Big(\sum_{\lambda_n^\phi\leq\Lambda^2} {\rm D}_n e^{-sM^2}-\sum_{\lambda_n^\phi\leq\Lambda^2}{\rm D}_n e^{-s\lambda_n^\phi}\Big)\,.
\end{equation}
Up to this point we have only used elementary identities. Now we want to establish the connection between the above calculation of $\delta S_{_{\rm EH}}^\phi$ performed using a Wilsonian UV hard cutoff $\Lambda$ and the usual calculation performed resorting to the proper-time method \cite{Schwinger:1951nm, Dunne:2004nc,Dunne:2012vv}. In this scheme one sums over the whole (infinite) spectrum of eigenvalues and regularizes UV-divergences introducing a lower integration bound $1/\Lambda_{\rm pt}^2$ in the proper-time integral (see footnote \ref{ptfootnote}). To establish the aforementioned connection from\,\eqref{ptidentity2}, we have then to send in the right-hand side $\Lambda\to\infty$ while keeping $\Lambda_{\rm pt}$ finite, and see if we recover the proper-time definition. In other words, we have to perform the replacement

{\small\begin{align}\label{ptidentity3}
		\lim_{\Lambda_{\rm pt}\to\infty}&\int_{1/\Lambda_{\rm pt}^2}^{\infty}\frac{\dd s}{s}\Big(\sum_{\lambda_n^\phi\leq\Lambda^2} {\rm D}_ne^{-sM^2}-\sum_{\lambda_n^\phi\leq\Lambda^2} {\rm D}_ne^{-s\lambda_n^\phi}\Big)\nn \\
		\longrightarrow &\int_{1/\Lambda_{\rm pt}^2}^{\infty}\frac{\dd s}{s}\lim_{\Lambda\to\infty}\Big(\sum_{\lambda_n^\phi\leq\Lambda^2} {\rm D}_n e^{-sM^2}-\sum_{\lambda_n^\phi\leq\Lambda^2}{\rm D}_n e^{-s\lambda_n^\phi}\Big)\nonumber\\
		=-&\int_{1/\Lambda_{\rm pt}^2}^{\infty}\frac{\dd s}{s}\sum_{n}^{\infty}{\rm D}_n e^{-s\lambda_n^\phi}\,+\,\int_{1/\Lambda_{\rm pt}^2}^{\infty}\frac{\dd s}{s}\lim_{\Lambda\to\infty}\Big(e^{-sM^2}\sum_{\lambda_n^\phi\leq\Lambda^2}{\rm D}_n\Big)\,. 
\end{align}}

\noindent
The first term in the third line of the above equation is easily recognized as the usual {\it definition} of the functional determinant within the proper-time regularization. Therefore, in order for the third line of\,\eqref{ptidentity3} to coincide with the proper-time definition of the fluctuation determinant, the second term in this line has to vanish. For a generic scale $M \neq \Lambda$, this term contains field dependent $\Lambda$-divergences that cannot be discarded tout-court. Instead, for $M=\Lambda$ such term vanishes thanks to the presence of the factor $e^{-s\Lambda^2}$, and the proper-time result is recovered.

\section{Conclusions and outlooks}
\label{Conclusions}

In the present work we have addressed a question which, in our opinion, lies at the
interface between two fundamental aspects of modern quantum field theory. On the one
hand, according to Wilson's lesson, a QFT has to be regarded, in general,
as an effective field theory endowed with a physical UV cutoff $\Lambda$, the
scale above which the theory has to be replaced by its UV completion. On the
other hand, in gauge theories the naive use of a hard cutoff is known to lead to
violations of gauge invariance, as happens for instance in the one-loop calculation of the
vacuum polarization tensor in QED, where a quadratically divergent longitudinal term
is generated once a hard cutoff is imposed on the loop momentum. This has led to the
common view that, in gauge theories, hard cutoffs should be abandoned in favour of
regularization schemes that preserve gauge invariance.

The aim of the present paper was to show that this conclusion is not unavoidable. We have
considered the Euler-Heisenberg correction to the free Maxwell action as a clean and
explicit setting where the issue can be studied in detail. The central point of our analysis
is that the Wilsonian cutoff should be implemented as a hard spectral cutoff on the eigenvalues of the gauge-covariant
fluctuation operator. In scalar QED this amounts to retaining only the eigenvalues $\lambda$ of
\,$-D^2+m^2$\, such that $\lambda\leq \Lambda^2$, while in fermionic QED the
corresponding prescription is implemented on the squared modulus of the eigenvalues of
$\slashed{D}+m$. Since the spectrum of these operators is gauge invariant, the resulting one-loop determinant is gauge invariant from the outset.

For both scalar and fermionic QED we have carried out the calculation explicitly in a
constant magnetic (electric) background. In the scalar case, the spectrum reduces to the
standard Landau-level spectrum of the operator $-D^2+m^2$, with the physical
cutoff $\Lambda$ selecting a finite number of levels and, for each of them, a finite domain in the
remaining continuous momenta. In the fermionic case, the corresponding eigenvalue
problem for the Euclidean Dirac operator was solved by taking into account the pairing of
complex-conjugate eigenvalues. In both cases, the
Euler-Heisenberg Lagrangian was obtained by performing the finite spectral sum and the
momentum integrals before expanding in the regime
$m^2,eB\ll \Lambda^2$.

The result of this procedure is twofold. First, the leading terms reproduce the usual
Euler-Heisenberg effective Lagrangian obtained with proper-time regularization, up to vacuum-energy contributions and after the standard analytic continuation to Lorentzian signature. In particular, the logarithmic
terms responsible for the one-loop running of the electric charge are correctly recovered.
Second, the hard spectral cutoff produces terms which are
absent in the proper-time result. These terms depend on the fractional part of
the maximal number of Landau levels included below the cutoff and are therefore periodic
functions of $1/eB$. When $eB\ll \Lambda^2$ they are
suppressed by powers of $eB/\Lambda^2$ and the usual (proper-time) result is recovered (further comments on these terms are at the end of the section). 

Another important point discussed in the present paper concerns the scale which makes the argument of the logarithm of
the fluctuation determinant dimensionless. Starting from the phase
space path integral, we argued that this scale is not an arbitrary fixed scale $M$, but it is
precisely the Wilsonian UV physical cutoff $\Lambda$. This identification is not a matter of
convention. We have shown that, if one keeps $M\neq \Lambda$, spurious field-dependent contributions appear in the Euler-Heisenberg action.
In scalar QED these terms would even modify the coefficient and the sign of the one-loop
beta function. Moreover, in scalar as well as in fermionic QED, spurious large-amplitude oscillatory terms are generated. Conversely, for $M=\Lambda$, all these unphysical contributions are absent. We have also shown that $M=\Lambda$ allows to establish a connection between our calculation of the fluctuation determinant with the Wilsonian hard cutoff $\Lambda$ and its definition within the framework of proper-time regularization.

The results of the present work provide a concrete example in which a hard cutoff,
endowed with a direct physical meaning, can be introduced in a gauge theory without
spoiling gauge invariance at the quantum level. In this sense, they represent a first step
towards a formulation of the Wilsonian renormalization group in gauge theories closer in
spirit to the Wegner-Houghton construction \cite{Wegner:1972ih}, where the integration of modes is performed
with a sharp cutoff. In fact, we suggest that in gauge
theories the notion of \vv mode'' should be tied to the spectrum of gauge-covariant
operators rather than to the Fourier momentum of the free theory.

Several directions deserve further investigation. First, it would be important to clarify
the relation between the hard spectral cutoff used here and lattice gauge theory. In the
lattice formulation, gauge invariance is implemented at the discretized level through link
variables rather than through the gauge potentials themselves. In perturbation theory this
leads, for example, to additional contributions such as seagull diagrams, which cancel the
quadratically divergent gauge-noninvariant terms produced by a naive momentum cutoff
in the vacuum polarization tensor. Here, gauge-noninvariant terms are absent due to the use a hard cutoff over the eigenvalues. 

A closely related problem is the comparison between the continuum spectral determinant
studied in this work and the determinant obtained from lattice QED in a constant magnetic
background, where the relevant spectral problem is connected with the Hofstadter
spectrum \cite{Hofstadter:1976, Janecek:2013}. Such a comparison would allow one to understand in detail how the Landau
level structure emerges from a microscopic gauge-invariant lattice regularization, and how
the periodic terms found here are represented in the lattice theory.

Another natural development is the construction of an exact or approximate
Wegner-Houghton-type renormalization group equation \cite{Wegner:1972ih} for QED, and more generally for
gauge theories. The calculations and results of the present work suggest that such a formulation is not
excluded by gauge invariance, provided the cutoff is imposed spectrally.

Finally, a comment on the terms periodic in $1/e B$ found in both scalar QED (forth line of\,\eqref{scalarresult}) and fermionic QED (fourth line of\,\eqref{fermionresult}). These terms
deserve further study both from a theoretical and a phenomenological point of view. They
originate from the discreteness of the spectrum below a sharp physical cutoff.  In this respect,
they are reminiscent of the oscillatory contributions that appear in phenomena such as
de Haas-van Alphen oscillations \cite{deHaasvanAlphen:1930, Shoenberg:1984}. Whether in gauge theories analogous effects can have observable
consequences in situations where the background field probes scales not much smaller
than the UV physical cutoff $\Lambda$ is an open question.

\section*{Acknowledgments}

We would like to thank D.\,Zappalà for useful discussions. The work of VB, FC, RG and AP is carried out within the INFN project  QGSKY.

\appendix

\section{Eigenvalue problem for $\slashed D+m$. Constant magnetic background}
\label{app1}

In this Appendix, we solve the eigenvalue problem for the (Euclidean) Dirac operator $\slashed D+m$ for a constant magnetic field background $B$. For the gauge field $A_\mu$ we take
\be \label{Amu}
A_0=A_2=A_3=0 \, , \qquad A_1=-B x_2  ,
\ee
so that the eigenvalue problem reads
\begin{equation}\label{eigenprob}
	(\gamma^\mu\partial_\mu-ieB\gamma^1x_2+m)\psi(x)=\lambda\psi(x)\,.
\end{equation}
We now write $\psi$ as
\begin{equation}\label{twocomp}
	\psi(x)=\begin{pmatrix}
		\varphi(x) \\
		\chi(x)
	\end{pmatrix}\,,
\end{equation}
where $\varphi(x)$ and $\chi(x)$ are two-component spinors,
\begin{equation}
	\varphi(x)=\begin{pmatrix}
		\varphi_1(x) \\
		\varphi_2(x)
	\end{pmatrix}\qquad,\qquad \chi(x)=\begin{pmatrix}
		\chi_1(x) \\
		\chi_2(x)
	\end{pmatrix}\,.
\end{equation}
Inserting\,\eqref{gammamatrices} and\,\eqref{twocomp} in\,\eqref{eigenprob} we get
\begin{align}
	C_1\chi(x)\equiv(\partial_0-i\sigma^j\partial_j-e B x_2\sigma^1)\chi(x)&=i\tilde\lambda\,\,\varphi(x)\label{eig1}\\
	C_2\varphi(x)\equiv(\partial_0+i\sigma^j\partial_j+e B x_2\sigma^1)\varphi(x)&=i\tilde\lambda\,\,\chi(x)\,,\label{eig2}
\end{align}
where $\tilde\lambda$ is defined as $\tilde\lambda\equiv -i(\lambda-m)$. Being $i\tilde\lambda$ the eigenvalues of the anti-hermitian operator $\slashed D$, $\tilde \lambda$ are real numbers. The system\,\eqref{eig1}-\eqref{eig2} can be solved in two equivalent ways. Either using first\,\eqref{eig2}, writing $\chi=(i\tilde\lambda)^{-1}C_2\varphi$ and inserting this latter relation in\,\eqref{eig1}, or using first\,\eqref{eig1}, writing $\varphi=(i\tilde\lambda)^{-1}C_1\chi$ and inserting this relation in\,\eqref{eig2}. 

Let us insert\,\eqref{eig2} in\,\eqref{eig1}, i.e.\,\,we write $\chi=(i\tilde\lambda)^{-1}C_2\varphi$. After some trivial algebra we get
\begin{align}
	[(-\partial_0^2-\partial_3^2)+(-\partial_1^2-\partial_2^2+2i e B x_2\partial_1+e^2 B^2 x_2^2)+e B\sigma^3]\varphi(x)=\tilde\lambda^2\,\varphi(x)\,,
\end{align}
that corresponds to the system 
\begin{align}
		&(-\partial_0^2-\partial_3^2-\partial_1^2-\partial_2^2+2i e B x_2\partial_1+e^2 B^2 x_2^2)\varphi_1=(\tilde\lambda^2-e B)\varphi_1\label{phi1}\\
		&(-\partial_0^2-\partial_3^2-\partial_1^2-\partial_2^2+2i e B x_2\partial_1+e^2 B^2 x_2^2)\varphi_2=(\tilde\lambda^2+e B)\varphi_2\label{phi2}\,.
\end{align}
In the left hand side of both\,\eqref{phi1} and\,\eqref{phi2} we recognize the eigenvalue problem for a free (non-relativistic) particle of mass $1/2$ in the $(x_0,x_3)$ plane and the Landau-level problem with frequency $2e B$ in the $(x_1,x_2)$ plane. For $(\tilde\lambda^2-e B)$ and $(\tilde\lambda^2+e B)$ we then have
\begin{align}
	(\tilde\lambda^\psi)^2\pm e B&=k_0^2+k_3^2+2 e B \Big(n+\frac12\Big)\,,\quad n=0,1,\dots\,,\,\,\,k_i=\frac{2\pi n_i}{L}\,,\,n_i\in\mathbb{Z}\,.
\end{align}
Therefore, for $\tilde\lambda$ we have
\begin{equation}\label{eigvalues}
	\tilde\lambda_{n,k_0,k_3}^\psi=\pm\sqrt{k_0^2+k_3^2+2 e B n}\,,\quad n=0,1,\dots\,,\,\,\,k_i=\frac{2\pi n_i}{L}\,,\,n_i\in\mathbb{Z}\,,
\end{equation}
and the eigenvalues $\lambda$ are (see comments below\,\eqref{eig2})
\begin{align}\label{eigvalues2}
	\lambda^\psi_{n,k_0,k_3}&=m+ i\sqrt{k_0^2+k_3^2+2 e B n}\,,\quad n=0,1,\dots\,,\,\,\,k_i=\frac{2\pi n_i}{L}\,,\,n_i\in\mathbb{Z}\,,\\
	(\lambda^\psi_{n,k_0,k_3})^*&=m- i\sqrt{k_0^2+k_3^2+2 e B n}\,,\quad n=0,1,\dots\,,\,\,\,k_i=\frac{2\pi n_i}{L}\,,\,n_i\in\mathbb{Z}\label{eigvalues4}\,.
\end{align}

\vskip 3pt

From now on we will use the notation $k^2 \equiv k_0^2+k_3^2$. To obtain now the  eigenfunctions corresponding to the eigenvalues \eqref{eigvalues3ferm}, we treat separately the case $n=0$ and the case $n\neq0$. Let us begin from the latter.

\vskip 8pt

\noindent
{\it Case $n\neq0$} 

\vskip 3pt

For any given $n$, we have two complex conjugate eigenvalues of $\slashed D+m$, namely $\lambda_{n,k_0,k_3}^\psi=m+i\tilde\lambda_{n,k_0,k_3}^\psi\equiv m+ i\sqrt{k^2+2 e B n}$ and its complex conjugate (see\,\eqref{eigvalues2} and\,\eqref{eigvalues4}). We will see that the eigenfunctions associated to $-\tilde\lambda_{n,k_0,k_3}^\psi$ are obtained applying the matrix $\gamma_5$ to the eigenfunctions associated to $\tilde\lambda_{n,k_0,k_3}^\psi$ (this is due to the fact that $\gamma_5$ and $\gamma^\mu$ anticommute). Let us write $\chi=(i\tilde\lambda)^{-1}C_2\varphi$, take $\tilde\lambda=\tilde\lambda_{n,k_0,k_3}^\psi \equiv \sqrt{k^2+2 e B n}$ and go back to the system \,\eqref{phi1}-\eqref{phi2}. To obtain the linearly independent eigenfunctions corresponding to $\lambda_{n,k_0,k_3}^\psi$\, we search for solutions of the form
\begin{equation}
	\varphi^{(1)}(x)=\begin{pmatrix}
		\varphi_1(x) \\
		0
	\end{pmatrix}\qquad\text{and}\qquad \varphi^{(2)}(x)=\begin{pmatrix}
		0 \\
		\varphi_2(x)
	\end{pmatrix}\,.
\end{equation}
This gives rise to the eigenfunctions of $\slashed D+m$ (up to a normalization constant)
{\small\begin{align}
	\big(\psi^{\text{\tiny$(1)$}}_{\lambda}\big)_{n,k_0,k_3;p_1}=\begin{pmatrix}
		\varphi_1(x) \\
		0\\
		(i\tilde\lambda_{n,k_0,k_3}^\psi)^{-1}C_2\begin{pmatrix}
			\varphi_1(x) \\
			0
		\end{pmatrix}
	\end{pmatrix}\,\,,\,\,	\big(\psi^{\text{\tiny$(2)$}}_{\lambda}\big)_{n,k_0,k_3;p_1}=\begin{pmatrix}
		0\\
		\varphi_2(x)\\
		(i\tilde\lambda_{n,k_0,k_3}^\psi)^{-1}C_2\begin{pmatrix}
			0\\
			\varphi_2(x) 
		\end{pmatrix}
	\end{pmatrix}\,.\label{eigfunctions1}
\end{align}}

\noindent
Resorting again to the well-known solution of the Landau-level problem with frequency $2eB$, we see that (see\,\eqref{eigscalar} and comments therein)
\begin{align}
	&\varphi_{1} (x) = \phi_{n-1,k_0,k_3;p_1}\label{solphi1} \\\nonumber\\
	&\varphi_{2} (x) =\phi_{n,k_0,k_3;p_1}\label{solphi2}  \, .
\end{align}
Since
\begin{equation}\label{pi}
	p_1^{\rm min}\equiv-\frac{e B}{2}L\leq p_1\leq\frac{e B}{2}L\equiv p_1^{\rm max}\,,
\end{equation}
the number of linearly independent eigenfunctions of type $\big(\psi^{\text{\tiny$(1)$}}_{\lambda}\big)_{n,k_0,k_3;p_1}$ (of type $\big(\psi^{\text{\tiny$(2)$}}_{\lambda}\big)_{n,k_0,k_3;p_1}$) associated to the eigenvalue\, $\lambda_{n,k_0,k_3}^\psi=m+i\sqrt{k^2+2 e B n\,}$\, is
\begin{equation}\label{deg1}
	\frac{L}{2\pi}(p_1^{\rm max}-p_1^{\rm min})=\frac{e B}{2\pi} L^2\,.
\end{equation}
The eigenvalue $\lambda_{n,k_0,k_3}^\psi$ has then degeneracy
\begin{equation}\label{deg3}
	D_{n,k_0,k_3}^\psi=\frac{e B}{\pi} L^2\,.
\end{equation}
The final expression of $\big(\psi^{\text{\tiny$(1)$}}_{\lambda}\big)_{n,k_0,k_3;p_1}$ and $\big(\psi^{\text{\tiny$(2)$}}_{\lambda}\big)_{n,k_0,k_3;p_1}$ is obtained once we calculate $C_2\varphi^{(1)}$ and $C_2\varphi^{(2)}$ in\,\eqref{eigfunctions1}. To this end, we now write explicitly the operator $C_2$
\begin{align}
	\text{\small$C_2=\partial_0+i\sigma_j\partial_j+e B x_2\sigma^1=\begin{pmatrix}
			\partial_0+i\partial_3 & \partial_2+i\partial_1+eBx_2\\
			-\partial_2+i\partial_1+eBx_2 & \partial_0-i\partial_3 
		\end{pmatrix}$}\,.
\end{align}
Since $C_2$ acts on $\varphi^{(1)}$ and $\varphi^{(2)}$, which depend on $x_0$, $x_1$ and $x_3$ only through the plane waves $e^{i k_0 x_0}$, $e^{i p_1 x_1}$ and $e^{i k_3 x_3}$, we can replace $\partial_0$ with $i k_0$, $\partial_1$ with $i p_1$ and  $\partial_3$ with $i k_3$, thus getting
\begin{align}
	\text{\small$C_2=\begin{pmatrix}
			i k_0-k_3 & \partial_2-p_1+eBx_2\\
			-\partial_2-p_1+eBx_2 & i k_0+k_3
		\end{pmatrix}\equiv\begin{pmatrix}
			i k_0-k_3 & \tilde\partial_2+eB\tilde x_2\\
			-\tilde\partial_2+eB\tilde x_2 &i k_0+k_3
		\end{pmatrix}$}\,,
\end{align}
where we have defined $\tilde x_2\equiv x_2-\frac{p_1}{eB}$ and $\tilde\partial_2\equiv\pdv{}{\tilde x_2}$. We can now write $C_2$ in terms of the creation and annihilation operators 
\begin{align}
	b^{\dagger}&=-\frac{1}{\sqrt{2e B}} (\tilde \partial_2-e B\tilde x_2 )\qquad,\qquad b=\frac{1}{\sqrt{2e B}} (\tilde \partial_2+e B\tilde x_2 )\label{cranH}
\end{align}
as
\begin{align}
	\text{\small$C_2=\begin{pmatrix}
			i k_0-k_3 & \sqrt{2e B\,}\,b\\
			\sqrt{2e B\,}\,b^{\dagger} & i k_0+k_3
		\end{pmatrix}$}\,.
\end{align}
We then get
\begin{align}
	C_2\varphi^{(1)}&=\begin{pmatrix}
		(ik_0-k_3)\phi_{n-1,k_0,k_3;p_1}\\
		\sqrt{2e B n\,}\,\phi_{n,k_0,k_3;p_1}
	\end{pmatrix}\\\nonumber\\
	C_2\varphi^{(2)}&=\begin{pmatrix}
		\sqrt{2e B n\,}\,\phi_{n-1,k_0,k_3;p_1}\\
		(ik_0+k_3)\phi_{n,k_0,k_3;p_1}
	\end{pmatrix}\,,
\end{align}
and from\,\eqref{eigfunctions1} we have
\begin{align}
	\big(\psi^{\text{\tiny$(1)$}}_{\lambda}\big)_{n,k_0,k_3;p_1}&=\begin{pmatrix}
		\phi_{n-1,k_0,k_3;p_1} \\
		0\\
		\frac{k_0+ik_3}{\sqrt{k^2+2e Bn}}\phi_{n-1,k_0,k_3;p_1}\\
		\frac{-i\sqrt{2e B n\,}}{\sqrt{k^2+2eBn}}\,\phi_{n,k_0,k_3;p_1}
	\end{pmatrix}\label{eigphi1}\\	\big(\psi^{\text{\tiny$(2)$}}_{\lambda}\big)_{n,k_0,k_3;p_1}&=\begin{pmatrix}
		0\\
		\phi_{n,k_0,k_3;p_1}\\
		\frac{-i\sqrt{2e B n\,}}{\sqrt{k^2+2eBn}}\,\phi_{n-1,k_0,k_3;p_1}\\
		\frac{k_0-ik_3}{\sqrt{k^2+2eBn}}\phi_{n,k_0,k_3;p_1}
	\end{pmatrix}\label{eigphi2}\,.
\end{align}
Finally, the eigenfunctions $\big(\psi^{\text{\tiny$(1)$}}_{\lambda^*}\big)_{n,k_0,k_3;p_1}$ and $\big(\psi^{\text{\tiny$(2)$}}_{\lambda^*}\big)_{n,k_0,k_3;p_1}$ associated to the complex conjugate eigenvalue $(\lambda_{n,k_0,k_3}^\psi)^*$ (see\,\eqref{eigvalues4}) are obtained from\,\eqref{eigphi1} and\,\eqref{eigphi2} through the replacement $\tilde\lambda_{n,k_0,k_3}^\psi=\sqrt{k^2+2eBn}\to-\tilde\lambda_{n,k_0,k_3}^\psi$. 
Using\,\eqref{gamma5}, we see that (as anticipated) $\big(\psi^{\text{\tiny$(1)$}}_{\lambda^*}\big)_{n,k_0,k_3;p_1}=-\gamma_5 \big(\psi^{\text{\tiny$(1)$}}_{\lambda}\big)_{n,k_0,k_3;p_1}$ and $\big(\psi^{\text{\tiny$(2)$}}_{\lambda^*}\big)_{n,k_0,k_3;p_1}=-\gamma_5 \big(\psi^{\text{\tiny$(2)$}}_{\lambda}\big)_{n,k_0,k_3;p_1}$. Therefore, the eigenvalue $(\lambda_{n,k_0,k_3}^\psi)^*$ has the same degeneracy $D_{n,k_0,k_3}^\psi$ (see\,\eqref{deg3}) as $\lambda_{n,k_0,k_3}^\psi$.

\vskip 8pt

\noindent
{\it Case $n=0$} 

\vskip 3pt

Let us consider now the case $ \tilde\lambda_{0,k_0,k_3}^\psi=\sqrt{k_0^2+k_3^2}=k$ (that corresponds to the eigenvalues $\lambda_{0,k_0,k_3}^\psi=m+ i k$ \,and\, $\lambda^*_{0,k_0,k_3}=m-i k$ \,of\, $\slashed D+m$). From\,\eqref{phi1} and\,\eqref{phi2} we see that solutions of the kind $\varphi^{(1)}$ are not allowed (since they are not normalizable), and that only solutions of the kind $\varphi^{(2)}$ are acceptable. The linearly independent eigenfunctions corresponding to $\lambda_{0,k_0,k_3}^\psi$ and $(\lambda_{0,k_0,k_3}^\psi)^*$ are
\begin{equation}
		\big(\psi^{\text{\tiny$(2)$}}_{\lambda}\big)_{0,k_0,k_3;p_1}=\begin{pmatrix}
			0\\
			\phi_{0,k_0,k_3;p_1}\\
			0\\
			\frac{k_0-ik_3}{k}\,\phi_{0,k_0,k_3;p_1}
		\end{pmatrix}\label{eigphiq1}
\end{equation}
and
\begin{equation}
	\,\big(\psi^{\text{\tiny$(2)$}}_{\lambda^*}\big)_{0,k_0,k_3;p_1}=\begin{pmatrix}
		0\\
		\phi_{0,k_0,k_3;p_1}\\
		0\\
		\frac{i k_3-k_0}{k}\,\phi_{0,k_0,k_3;p_1}
	\end{pmatrix}\,,\label{eigphiq2}
\end{equation}
respectively. The eigenvalues $\lambda_{0,k_0,k_3}^\psi$ and $(\lambda_{0,k_0,k_3}^\psi)^*$ have then both degeneracy $D_{0,k_0,k_3}^\psi=\frac{e B}{2\pi} L^2$ (see comments below\,\eqref{solphi2}).

To summarize, we have found the complete set of orthogonal eigenfunctions (see\,\eqref{eigphi1}, \eqref{eigphi2}, \eqref{eigphiq1}, \eqref{eigphiq2})
\begin{equation}\label{basis}
	\big(\psi^{\text{\tiny$(1)$}}_{\lambda}\big)_{n,k_0,k_3;p_1},\big(\psi^{\text{\tiny$(1)$}}_{\lambda^*}\big)_{n,k_0,k_3;p_1},\big(\psi^{\text{\tiny$(2)$}}_{\lambda}\big)_{n,k_0,k_3;p_1},\big(\psi^{\text{\tiny$(2)$}}_{\lambda^*}\big)_{n,k_0,k_3;p_1},\big(\psi^{\text{\tiny$(2)$}}_{\lambda}\big)_{0,k_0,k_3;p_1},\big(\psi^{\text{\tiny$(2)$}}_{\lambda^*}\big)_{0,k_0,k_3;p_1}\,,
\end{equation}
that spans the space of the field $\psi$. Similarly, the space of $\bar\psi$ is spanned by the set
\begin{equation}\label{basis2}
	\big(\psi^{\text{\tiny$(1)$}}_{\lambda}\big)^\dagger_{n,k_0,k_3;p_1},\big(\psi^{\text{\tiny$(1)$}}_{\lambda^*}\big)^\dagger_{n,k_0,k_3;p_1},\big(\psi^{\text{\tiny$(2)$}}_{\lambda}\big)^\dagger_{n,k_0,k_3;p_1},\big(\psi^{\text{\tiny$(2)$}}_{\lambda^*}\big)^\dagger_{n,k_0,k_3;p_1},\big(\psi^{\text{\tiny$(2)$}}_{\lambda}\big)^\dagger_{0,k_0,k_3;p_1},\big(\psi^{\text{\tiny$(2)$}}_{\lambda^*}\big)^\dagger_{0,k_0,k_3;p_1}\,.
\end{equation}

\end{document}